\documentclass[10pt]{article}
\usepackage{amssymb}  
\usepackage{bm}
\usepackage{amsfonts}
\usepackage{amsmath}
\usepackage{authblk}
 
\newtheorem{prop}{Proposition}[section] 
 
\newtheorem{rem}{Remark}[section]

\def\cuuno{\mathop{q_1}\limits}
\def\cuell{\mathop{q_\ell}\limits}
\def\cu{\mathop{\bf Q}\limits}
\def\p{\mathop{P}\limits}
\def\ics{\mathop{\bf X}\limits}
\def\cuk{\mathop{q_k}\limits}
\def\cub{\mathop{\bf q}\limits}
\def\cubtil{\mathop{\bar {\bf q}}\limits}

\title{An extended Lagrangian formalism}
\author{F.~Talamucci}
\affil{DIMAI, Dipartimento di Matematica e Informatica ``Ulisse Dini'',\\
Universit\`a degli Studi di Firenze, Italy\footnote{
corresponding author e-mail: federico.talamucci@unifi.it}}
\date{}

\begin{document}
\bibliographystyle{plain}

\setcounter{equation}{0}
\setcounter{ese}{0}
\setcounter{eserc}{0}
\setcounter{teo}{0}               
\setcounter{corol}{0}
\setcounter{propr}{0}

\maketitle

\vspace{.5truecm}

\noindent
{\bf Keywords:}

\vspace{.5truecm}

\vspace{.5truecm}

\noindent
{\bf Abstract}. A simple formal procedure makes the main properties of the lagrangian binomial $\nabla_{\bf q}{\cal L}-\frac{d}{dt} \nabla_{\dot {\bf q}}{\cal L}$ extendable to functions depending to any kind of order of the time--derivatives of the lagrangian coordinates ${\bf q}$.
Such a broadly formulated binomial can provide the lagrangian components, in the classical sense of the Newton's law, for a quite general class of forces.  

\noindent
At the same time, the generalized equations of motions recover some of the classical alternative formulations of the Lagrangian equations.

\vspace{.5truecm}

\section{Introduction}

\noindent
The starting point of our investigation is a set of covariant expressions, which are listed all togheter in order to draw attention to their common origin. The differential operators we are going to specify act on functions depending on ${\bf q}$, ${\dot {\bf q}}$, $\dots$ at any fixed order of derivative. 
Thus the question arises, about which role plays such operators within the framework of the formulation of the dynamical laws for discrete systems of material points. The additional aspect, with respect to the standard theory, is evidently the presence of higher order derivatives of the lagrangian coordinates.

\noindent
As we will see, the main point we are going to discuss is to examine 
a generalized lagrangian binomial in order to keep the standard properties of the binomial concerning the standard case and at the same time to ascribe a physical meaning so that, just like the standard case, one can make us of the generalized binomial for the proper equations of motion.

\noindent
The notation we use shows $\cuk^{h\cdot}$ standing for $q_k$ with $h$ dots: $\cuk^{\overbrace{{\cdot \cdot \;\cdot \cdot}}^{h}}$, that is the $h$-th derivative with respect to $t$ $\dfrac{d^h q_k}{dt^h}$; the same for $\cub^{h\cdot}=(\cuuno^{h\cdot}, \dots, \cuell^{h\cdot})$.
Unless stated otherwise, the notation is valid also for $s=0$, in the sense of the $0$--derivative  $\cuk^{0\cdot}=q_k$, $\dfrac{d^0 q_k}{dt^0}=q_k$. 

\noindent
For a given real--valued function ${\cal Y}_r({\bf q}, {\dot {\bf q}}, \dots, \cub^{r\cdot},  \cub^{(r+1)\cdot}, t)$, $r\geq 0$ and a given index $h=0,1,\dots, r+1$, consider the $r+2$ operators
\begin{equation}
\label{opcomp}
\sum\limits_{j=0}^h (-1)^j\left(\begin{array}{c} r+1-h+j \\ r+1 -h \end{array}\right)\dfrac{d^j}{dt^j}
\nabla_{\cub^{ (r+1-h+j)\cdot}}{\cal Y}_r
\end{equation}
which transform ${\cal Y}_r$ into a $\ell$--vector function. For a fixed value of $h$, the corresponding expression (\ref{opcomp}) contain  $h+1$ terms, which are worthy to be written explicitly, for greater clarity:

\begin{equation}
\label{opercovar}
\begin{array}{ll}
h=r+1: &\nabla_{\bf q}{\cal Y}_r - \frac{d}{dt} \nabla_{\dot {\bf q}}{\cal Y}_r+ \frac{d^2}{dt^2}
\nabla_{\cub^{..}}{\cal Y}_r+ \dots  +(-1)^{r+1}\dfrac{d^{r+1}}{dt^{r+1}}
\nabla_{\cub^{(r+1)\cdot}}{\cal Y}_r, \\
h=r: &\nabla_{\dot {\bf q}}{\cal Y}_r - 2\frac{d}{dt} \nabla_{\cub^{..}}{\cal Y}_r+ 3\frac{d^2}{dt^2}
\nabla_{\cub^{...}}{\cal Y}_r+ \dots +(-1)^r(r+1)\frac{d^r}{dt^r}
\nabla_{\cub^{(r+1) \cdot}}{\cal Y}_r,\\
& \dots \\
h: & 
\nabla_{\cub^{(r+1-h)\cdot}}{\cal Y}_r - (r+1-h+1)
\frac{d}{dt} \nabla_{\cub^{(r+1-h+1)\cdot}}{\cal Y}_r+ \dots \\
&\dots  +(-1)^j
\left(\begin{array}{c} r+1-h+j \\ r+1 -h \end{array}\right)
\frac{d^j}{dt^j}\nabla_{\cub^{(r+1-h+j)\cdot}}{\cal Y}_r+\dots  \\
& 
\dots +(-1)^h
\left(\begin{array}{c} r+1 \\ r+1 -h \end{array}\right)
\frac{d^h}{dt^h}\nabla_{\cub^{(r+1)\cdot}}{\cal Y}_r,
\\
& \dots \\
h=1: &\nabla_{\cub^{r\cdot}} {\cal Y}_r
-(r+1)\frac{d}{dt}\nabla_{\cub^{(r+1)\cdot}} {\cal Y}_r, \\
h=0: &\nabla_{\cub^{(r+1)\cdot}} {\cal Y}_r.
\end{array}
\end{equation}

\noindent
Let ${\bar {\bf q}}({\bf q},t)$ be a 
${\cal C}^\infty$--change of coordinates, with $det\,J_{\bf q}{\bar {\bf q}}\not =0$, where $\left(J_{\bf q}{\bar {\bf q}}\right)_{i,j}=\dfrac{\partial {\bar q}_i}{\partial q_j}$, $i,j=1,\dots, \ell$. 
For the given real--valued function ${\cal Y}_r({\bf q}, {\dot {\bf q}}, \dots, \cub^{r\cdot},  \cub^{(r+1)\cdot}, t)$ we denote by
$\widetilde {\cal Y}_r({\bf q},$ ${\dot {\bf q}},$ $\dots,$ $\cubtil^{r\cdot},$ $\cubtil^{(r+1)\cdot}, t)$ the function ${\cal Y}_r$ calculated in ${\bf q}({\bar {\bf q}}, t)$, ${\dot {\bf q}}({\bar {\bf q}}, {\dot {\bar {\bf q}}}, t)$, $\dots$,  $\cub^{r\cdot}({\bar {\bf q}}, {\dot {\bar {\bf q}}}, \dots, \cubtil^{r\cdot},t)$, 
$\cub^{(r+1)\cdot}({\bar {\bf q}},$ ${\dot {\bar {\bf q}}},$ $\dots, \cubtil^{(r+1)\cdot},t)$ (the existence of the inverse functions is guaranteed by the non singularity of $J_{\bf q}{\bar {\bf q}}$).

\noindent
The important feature of (\ref{opcomp}) is that each of the $r+2$ operators is covariant: the case $h=r+1$ is well known, since it arises when considering the equations originated by a Lagrangian function ${\cal Y}_r$, containing  derivatives at higher orders than the standard case. On the other hand, the case $h=0$ is also evident, considering that 
$$
\nabla_{\cub^{(r+1)\cdot}} {\cal Y}_r=J_{\bf q}^T{\bar {\bf q}}\nabla_{\cubtil^{(r+1)\cdot}}{\widetilde {\cal Y}}_r
$$
where the apex $T$ designates the transposition of the matrix.
The key point in order to proof the covariance for any $h$ is the property
\begin{equation}
\label{propdot}
J_{\cub^{k\cdot}}\cubtil^{(k+i)\cdot}=\left(\begin{array}{c} k+i \\ h \end{array}\right)
\frac{d^i}{dt^i}J_{\bf q}{\bar {\bf q}}, \quad i\geq 0
\end{equation}
which can be directly checked, or can be traced, for istance, in \cite{craig}.
The case $i=0$ shows the known property of cancelling dots in the jacobian matrices of the transformations between variables with the same number of dots.

\noindent
Rather than producing a general proof of covariance, we will deal directly with the binomial corresponding to the special case $h=1$ and we will associate such a case with concrete formulations of mechanical systems.
It must be said that our investigation is strongly based upon \cite{ming}, where equation $(7)$ of the Introduction 
inspired our proposal of generalization.

\section{Extended Lagrangian binomial} 

\noindent
We focus on the case $h=1$ with respect to (\ref{opcomp}) and we examine the operator 
\begin{equation}
\label{extlagr}
{\cal O}_r[{\cal Y}_r]=\nabla_{\cub^{r\cdot}} {\cal Y}_r
-(r+1)\frac{d}{dt}\nabla_{\cub^{(r+1)\cdot}} {\cal Y}_r
\end{equation}
which is in our mind the generalization of the Lagrangian binomial (corresponding to $r=0$) ${\cal O}_0[{\cal L}]=\nabla_{\bf q} {\cal L}
-\frac{d}{dt}\nabla_{\dot {\bf q}} {\cal L}$, for any ${\cal L}({\bf q}, {\dot {\bf q}},t)$.
\begin{prop} The operator (\ref{extlagr})is covariant.
\end{prop}
{\bf Proof.} 
We start from the calculation
\begin{eqnarray*}
\nabla_{\cub^{(r+1)\cdot}} {\cal Y}_r
&=&J_{\cub^{(r+1)\cdot}}^T\cubtil^{(r+1)\cdot}\nabla_{\cubtil^{(r+1)\cdot}}{\widetilde {\cal Y}}_r,\\
\nabla_{\cub^{r\cdot}} {\cal Y}_r
&=&J_{\cub^{r\cdot}}^T\cubtil^{r\cdot}\nabla_{\cubtil^{r\cdot}}{\widetilde {\cal Y}}_r+
J_{\cub^{r\cdot}}^T\cubtil^{(r+1)\cdot}\nabla_{\cubtil^{(r+1)\cdot}}{\widetilde {\cal Y}}_r
\end{eqnarray*}
and we make use of the property (\ref{propdot}) in the specific cases
$$
\begin{array}{ll}
J_{\cub^{(r+1)\cdot}}\cubtil^{(r+1)\cdot}=J_{\cub^{r\cdot}}\cubtil^{r\cdot}=J_{\bf q}{\bar {\bf q}}, &
J_{\cub^{r\cdot}}\cubtil^{(r+1)\cdot}=(r+1)\frac{d}{dt}J_{\bf q}{\bar {\bf q}}.
\end{array}
$$
Thus
$$
\begin{array}{l}
\nabla_{\cub^{r\cdot}} {\cal Y}_r-(r+1)
\frac{d}{dt}\nabla_{\cub^{(r+1)\cdot}} {\cal Y}_r\\
=J_{\bf q}^T{\bar {\bf q}}\nabla_{\cubtil^{r\cdot}}{\widetilde {\cal Y}}_r+(r+1)\frac{d}{dt}
(J_{\bf q}^T{\bar {\bf q}})\nabla_{\cubtil^{(r+1)\cdot}}{\widetilde {\cal Y}}_r
\\
-(r+1)J_{\bf q}^T{\bar {\bf q}}\frac{d}{dt}\nabla_{\cubtil^{(r+1)\cdot}}{\widetilde {\cal Y}}_r
-(r+1)\frac{d}{dt}(J_{\bf q}^T{\bar {\bf q}})\nabla_{\cubtil^{(r+1)\cdot}}{\widetilde {\cal Y}}_r
\\
=
J_{\bf q}^T{\bar {\bf q}}\left(
\nabla_{\cubtil^{r\cdot}}{\widetilde {\cal Y}}_r-(r+1)
\frac{d}{dt}\nabla_{\cubtil^{(r+1)\cdot}}{\widetilde {\cal Y}}_r\right)
\end{array}
$$
whence the covariance of (\ref{extlagr}). $\quad\square$

\noindent
Additional properties we need concern the effect of (\ref{extlagr}) on the derivatives of a given function: we 
start from the following 
\begin{prop} For any  ${\cal L}({\bf q}, {\dot {\bf q}}, t)$ it holds ${\cal O}_0[{\cal L}]={\cal O}_r\left[\frac{d^r{\cal L}}{dt^r}\right]$, that is
\begin{equation}
\label{ext1r}
\nabla_{\bf q}{\cal L}-\frac{d}{dt} \nabla_{\dot {\bf q}}{\cal L}=
\nabla_{\cub^{r\cdot}}\frac{d^r{\cal L}}{dt^r}-(r+1)\frac{d}{dt}
\left(\nabla_{\cub^{(r+1)\cdot}}\frac{d^r{\cal L}}{dt^r}\right).
\end{equation}
\end{prop}

{\bf Proof.} 
We first remark that 
$\frac{d^r{\cal L}}{dt^r}=\nabla_{\dot {\bf q}}{\cal L}\cdot \cub^{(r+1)\cdot}+{\bm \Phi}({\bf q}, {\dot {\bf q}}, \dots, \cub^{r\cdot},t)$
hence 
$\nabla_{\cub^{(r+1)\cdot}} \frac{d^r{\cal L}}{dt^r}=\nabla_{\dot {\bf q}}{\cal L}$
so that (\ref{ext1r}) is equivalent to 
\begin{equation}
\label{r}
\nabla_{\cub^{r\cdot}}\frac{d^r{\cal L}}{dt^r}
=\nabla_{\bf q}{\cal L}+r\frac{d}{dt}\nabla_{\dot {\bf q}}{\cal L}.
\end{equation}
For $r=1$ it is immediate to check that $\nabla_{\dot {\bf q}}\frac{d{\cal L}}{dt}
=\nabla_{\bf q}{\cal L}+\frac{d}{dt}\nabla_{\dot {\bf q}}{\cal L}$; for $r>1$ we continue by induction on $r$, assuming that (\ref{r}) holds. Let us take advantage of the property concerning the inversion between total derivative and gradient:
\begin{equation}
\label{prop2}
\nabla_{\cub^{n\cdot}} \frac{dP}{dt}- \frac{d}{dt}\nabla_{\cub^{n\cdot}}P=\nabla_{\cub^{(n-1)\cdot}} P
\end{equation}
which is valid for any $P({\bf q}, {\dot {\bf q}}, \dots, \cub^{k\cdot}, t)$ and any $n\geq 1$. 
By employing (\ref{prop2}) with $P=\frac{d^r{\cal L}}{dt^r}$ and $n=r+1$ one achieves
$$
\nabla_{\cub^{(r+1)\cdot}} \frac{d^{r+1}{\cal L}}{dt^{r+1}} =
\nabla_{\cub^{r\cdot}} \frac{d^r{\cal L}}{dt^r}+
\frac{d}{dt}\left(\nabla_{\cub^{(r+1)\cdot}}\frac{d^r{\cal L}}{dt^r}\right)=
\nabla_{\cub^{r\cdot}} \frac{d^r{\cal L}}{dt^r}
+\frac{d}{dt}\nabla_{\dot {\bf q}}{\cal L}
$$
and, by virtue of the induction assumption (\ref{r}), 
$$
\nabla_{\cub^{(r+1)\cdot}} \frac{d^{r+1}{\cal L}}{dt^{r+1}} =
\nabla_{\bf q}{\cal L}+(r+1)\frac{d}{dt}\nabla_{\dot {\bf q}}{\cal L}
$$
hence (\ref{r}) holds also for $r+1$. $\quad\square$

\noindent
We render now relation (\ref{prop2}) more general by means of the following
\begin{prop}
For any function ${\cal U}({\bf q}, {\dot {\bf q}}, \dots, \cub^{s\cdot}, \cub^{(s+1)\cdot}, t)$, $s\geq 0$  and any index $r\geq s$ it is 
\begin{equation}
\label{prop13}
{\cal O}_r\left[ \frac{d^{r-s}{\cal U}}{dt^{r-s}}\right] = {\cal O}_s[{\cal U}].
\end{equation} 
\end{prop}
{\bf Proof.} 
It has to be checked that 
\begin{equation}
\label{extsr}
\nabla_{\cub^{r\cdot}}\frac{d^{r-s}{\cal U}}{dt^{r-s}}-(r+1)\frac{d}{dt}\left(
\nabla_{\cub^{(r+1)\cdot}}\frac{d^{r-s}{\cal U}}{dt^{r-s}}\right)
=\nabla_{\cub^{s\cdot}}{\,\cal U}-(s+1)\frac{d}{dt}\nabla_{\cub^{(s+1)\cdot}}{\cal U}.
\end{equation}

Since $\nabla_{\cub^{(r+1)\cdot}}\frac{d^{r-s}{\cal U}}{dt^{r-s}}=\nabla_{\cub^{(s+1)\cdot}}{\cal U}$, (\ref{extsr}) can be written as
\begin{equation}
\label{extsrequiv}
\nabla_{\cub^{r\cdot}}\frac{d^{r-s}{\cal U}}{dt^{r-s}}=(r-s)\frac{d}{dt}
\nabla_{\cub^{(s+1)\cdot}}{\cal U}+\nabla_{\cub^{s\cdot}}{\,\cal U}.
\end{equation}
We refer again to (\ref{prop2}) which is used $r-s$ times for $P=\frac{d^{r-s-1}{\cal U}}{dt^{r-s-1}}$,
$P=\frac{d^{r-s-2}{\cal U}}{dt^{r-s-2}}$, $\dots$, $P={\cal U}$ and $n=r$, $r-1$, $\dots$, $r-(r-s)+1=s+1$ respectively:
$$
\begin{array}{l}
\nabla_{\cub^{r\cdot}}\frac{d^{r-s}{\cal U}}{dt^{r-s}}
=
\nabla_{\cub^{r\cdot}} \left(\frac{d}{dt}
\frac{d^{r-s-1}{\cal U}}{dt^{r-s-1}}\right)
=\frac{d}{dt}\left(\nabla_{\cub^{r\cdot}}
\frac{d^{r-s-1}{\cal U}}{dt^{r-s-1}}\right)+
\nabla_{\cub^{(r-1)\cdot}} 
\frac{d^{r-s-1}{\cal U}}{dt^{r-s-1}}, \\
\\
\nabla_{\cub^{(r-1)\cdot}} 
\frac{d^{r-s-1}{\cal U}}{dt^{r-s-1}}=
\nabla_{\cub^{(r-1)\cdot}} \left(\frac{d}{dt}
\frac{d^{r-s-2}{\cal U}}{dt^{r-s-2}}\right)
=\frac{d}{dt}\left(\nabla_{\cub^{(r-1)\cdot}}
\frac{d^{r-s-2}{\cal U}}{dt^{r-s-2}}\right)+
\nabla_{\cub^{(r-2)\cdot}} \frac{d^{r-s-2}{\cal U}}{dt^{r-s-2}}, \\
\qquad \dots \quad \dots \quad \dots \\
\nabla_{\cub^{(s+1)\cdot}} \frac{d}{dt}{\cal U}=
\frac{d}{dt}\nabla_{\cub^{(s+1)\cdot}}{\cal U}+\nabla_{\cub^{s\cdot}}\,{\cal U}
\end{array}
$$
Since 
$$
\nabla_{\cub^{(r-1)\cdot}}
\left(\dfrac{d^{r-s-1}{\cal U}}{dt^{r-s-1}}\right)=
\nabla_{\cub^{(r-1)\cdot}}
\left(\dfrac{d^{r-s-2}{\cal U}}{dt^{r-s-2}}\right)=
\dots =
\nabla_{\cub^{(s+1)\cdot}}{\cal U},
$$ 
relation (\ref{extsrequiv}) easily follows.
We remark that (\ref{ext1r}) is (\ref{extsr}) whenever $s=0$.  $\quad\square$

\section{Extending kinetic energy and potential forces}

\noindent
In order to place (\ref{propdot}) and the other properties discussed above in a proper context of a real physical system, let us put our problem in the context of a system of $N$ material points $(P_i, m_i)$, $i=1,\dots, N$ and write the cartesian coordinates as a function of the lagrangian coordinates according to ${\bf X}({\bf q},t)\in {\Bbb R}^{3N}$, ${\bf q}\in {\Bbb R}^\ell$. As can be seen, we will consider only holonomic systems.
Calling ${\bf Q}=(m_1{\dot P}_1, \dots, m_N {\dot P}_N)$ the ${3N}$--vector of the momenta, we define

\begin{equation}
\label{deftr}
{\cal T}_r=\frac{1}{2}\cu^{r\cdot}\cdot\ics^{(r+1)\cdot}.
\end{equation}
For $r=0$ (\ref{deftr}) is the standard kinetic energy ${\cal T}_0=\frac{1}{2}{\bf Q}\cdot {\dot {\bf X}}=
\frac{1}{2}\sum\limits_{i=1}^N m_i{\dot P}_i^2$.
A second case to be remarked is $r=1$, providing the \emph{acceleration energy} (see \cite{gant}):
\begin{equation}
\label{accen}
{\cal T}_1=\frac{1}{2} \cu^{.} \cdot \ics^{..}=\frac{1}{2}\sum\limits_{i=1}^N m_i {\p^{..}}^2_i.
\end{equation}
The well known relation between the lagrangian components of ${\bf Q}$ and the lagrangian binomial calculated for ${\cal T}_0$ can be expanded to higher derivatives by means of the following
\begin{prop}

\begin{equation}
\label{binlagrt}
\cu^{(r+1)\cdot}\cdot \frac{\partial {\bf X}}{\partial q_k}=\frac{d}{dt} 
\dfrac{\partial {\cal T}_r}{\partial \cuk^{(r+1)\cdot}}
-\frac{1}{r+1}\,\frac{\partial {\cal T}_r}{\partial \cuk^{r\cdot}}, \;\;k=1,\dots, \ell
\end{equation}
\end{prop}

\noindent
{\bf Proof.} It suffices to replicate the steps of the standard case $r=0$, by writing
$$
\cu^{(r+1)\cdot}\cdot \frac{\partial {\bf X}}{\partial q_k}=\dfrac{d}{dt}
\cu^{r\cdot}\cdot \frac{\partial {\bf X} }{\partial q_k}-\cu^{r\cdot}\cdot \frac{\partial {\dot {\bf X}}}{\partial q_k}, \;\;k=1,\dots, \ell.
$$
On the other hand, one has
$$
\begin{array}{ll}
\frac{\partial {\cal T}_r}{\partial \cuk^{(r+1)\cdot}}=
\cu^{r\cdot}\cdot \dfrac{\partial \ics^{(r+1)\cdot}}{\partial \cuk^{(r+1)\cdot}}=
\cu^{r\cdot}\cdot \dfrac{\partial {\bf X}}{\partial q_k}, &
\dfrac{\partial {\cal T}_r}{\partial \cuk^{r\cdot}}=
\cu^{r\cdot}\cdot \dfrac{\partial \ics^{(r+1)\cdot}}{\partial \cuk^{r\cdot}}
=(r+1)\cu^{r\cdot}\cdot \dfrac{\partial {\bf X}}{\partial q_k}
\end{array}
$$
where the second equalities in each of the two sequences are deduced from relations analogous to (\ref{propdot}). $\quad\square$

\begin{rem}
In the case $r=1$, the acceleration energy (\ref{accen}) verifies
$$
\dfrac{\partial {\cal T}_1}{\partial \cuk^{\cdot \cdot}}={\dot {\bf Q}}\cdot \dfrac{\partial {\bf X}}{\partial q_k}
$$
and the Newton's law makes us write the equations of motion in the form 
$$
\dfrac{\partial {\cal T}_1}{\partial \cuk^{\cdot \cdot}}={\cal F}_k, \quad k=1,\dots, \ell
$$
where ${\cal F}_k$ is the $k$--lagrangian component of the force. The just written equations are the Appell equations (see \cite{app}), in the special case of pseudovelocities matching with the generalized velocities.
\end{rem}

\noindent
Assume now that the system of forces ${\cal F}\in {\Bbb R}^{3N}$ exerting on $(P_1, \dots, P_N)$ admits a function ${\cal U}({\bf q}, {\dot {\bf q}}, \dots, \cub^{(s+1)\cdot}, t)$, $s\geq 0$, such that

\begin{equation}
\label{potgen}
{\cal F}\cdot \dfrac{\partial {\bf X}}{\partial q_k}=
\frac{\partial {\cal U}}{\partial \cuk^{s\cdot}}-(s+1)\frac{d}{dt}
\frac{\partial {\cal U}}{\partial \cuk^{(s+1)\cdot}}.
\end{equation}
The case $s=0$ corresponds to a generalized potential ${\cal U}({\bf q}, {\dot {\bf q}},t)$, as for instance the Lorentz force in an electromagnetic field or the couple Coriolis force--centrifugal force exerted by a reference frame in uniform rotation. Both circumstances of Lorentz force and non inertial forces fall in the case of linear dependence on the velocity
$$
{\cal F}({\bf q}, {\dot {\bf q}},t)=
\left(J_{\bf q}^T{\bm \alpha}-J_{\bf q}{\bm \alpha}\right)^T{\dot {\bf q}}+\nabla_{\bf q}\beta -\dfrac{\partial {\bm \alpha}}{\partial t}
$$
(with appropriate ${\bm \alpha}$ and $\beta$) combined with the generalized potential ${\cal U}_1({\bf q}, {\dot {\bf q}},t)={\bm \alpha}({\bf q},t)\cdot {\dot {\bf q}}+\beta ({\bf q}, t)$.

\noindent
In order to improve notations, if one introduces the matrix--vector product
$
J_{\bf q}^T{\bf X}{\bf w}=
\left(
\begin{array}{c}
{\bf w}\cdot \frac{\partial {\bf X}}{\partial q_1} \\
\dots 
\\
{\bf w}\cdot \frac{\partial {\bf X}}{\partial q_\ell}
\end{array}
\right)
$
providing the $\ell$ lagrangian components of any ${3n}$--vector ${\bf w}$, (\ref{binlagrt}) and (\ref{potgen}) can be written in terms of the operator (\ref{extlagr}) respectively as
\begin{equation}
\label{tuop}
J_{\bf q}^T{\bf X}\cu^{(r+1)\cdot}=-\frac{1}{r+1}{\cal O}_r[{\cal T}_r], \quad 
J_{\bf q}^T{\bf X}{\cal F}={\cal O}_s[{\cal U}].
\end{equation}
It is also worthy to remark that (\ref{binlagrt}) allows us to place in the context of (\ref{extlagr}) special forces proportional to the acceleration, or the rate of change of acceleration or even further derivatives with respect to time:
\begin{equation}
\label{fa}
{\cal F}_A={\Bbb L}\ics^{(h+1)\cdot}
\end{equation}
where $h\geq 1$ and ${\Bbb L}$ is a diagonal matrix with positive entries. Calculations very close to those concerning (\ref{binlagrt}) lead to
$$
{\Bbb L}\ics^{(h+1)\cdot}\cdot \dfrac{\partial {\bf X}}{\partial q_k}=\frac{d}{dt} 
\dfrac{\partial {\cal U}_A}{\partial \cuk^{h\cdot}}
-\frac{1}{h}\,\frac{\partial {\cal U}_A}{\partial \cuk^{(h-1)\cdot}}, \;\;k=1,\dots, \ell
$$
where 
${\cal U}_A=\dfrac{1}{2}{\Bbb L}\ics^{h\cdot}\cdot \ics^{h\cdot}$. According to the notation of (\ref{extlagr}) we can also write
\begin{equation}
\label{ul}
J_{\bf q}^T{\bf X} {\cal F}_A=-\frac{1}{h}{\cal O}_{h-1}[{\cal U}_A].
\end{equation}
For $h=1$ and ${\Bbb L}$ listing the masses of the points one recovers the inertial forces. A second 
circumstance to have in mind is the Abraham--Lorentz force
(due to the electromagnetic radiation on an accelerating charged particle), which can be modeled choosing $h=2$:
$$
{\cal F}_A={\Bbb L}\ics^{...}, \quad 
{\cal U}_A=\frac{1}{2} {\Bbb L}\ics^{..}\cdot \ics^{..}
$$
where the diagonal matrix ${\Bbb L}$ depends on the charge of the particle, the speed of light and on the electric and magnetic constants. The lagrangian components of the force verify in this case
$$
{\cal F}_A\cdot \frac{\partial {\bf X}}{\partial q_k}=\frac{d}{dt}
\frac{\partial {\cal U}_A}{\partial \cuk^{..}}-\frac{1}{2}
\frac{\partial {\cal U}_A}{\partial {\dot q}_k}.
$$

\section{Other types of forces}

\noindent
Let us make a very simple point: for any function $\gamma ({\bf q}, {\dot {\bf q}}, \dots, \cub^{i\cdot}, t)$ and any index $h\geq 1$ it is
\begin{equation}
\label{point}
{\cal O}_i[\gamma]=\nabla_{\cub^{i\cdot}}\gamma.
\end{equation}
This allows us to count the following cases in the formal stucture (\ref{extlagr}):

\begin{itemize}
\item[$(i)$] a system of forces ${\cal F}_R$ such that the lagrangian components verify
\begin{equation}
\label{rayr}
J_{\bf q}^T{\bf X}{\cal F}_R =-\nabla_{\cub^{\sigma\cdot}}{\cal R}({\bf q}, {\dot {\bf q}}, \dots, \cub^{\sigma \cdot}, t)
\end{equation}
for some function ${\cal R}$ and some positive integer $\sigma$;
\item[$(ii)$] 
a general force ${\cal G}({\bf q}, {\dot {\bf q}}, \dots, \cub^{(\varrho-1)\cdot}, t)$, $\varrho \geq 1$.
\end{itemize}
In the first case, by virtue of (\ref{point}) we have 
\begin{equation}
\label{complagrfr}
{\cal O}_\sigma [-{\cal R}] = J_{\bf q}^T{\bf X}{\cal F}_R.
\end{equation}
The signus $-$ is legitimized if we think of the case $\sigma=1$, producing the dissipation force ${\cal F}_R=-J_{\bf q}{\bf X}_\mu{\dot {\bf q}}$, where ${\bf X}_\mu=(\mu_1P_1, \dots, \mu_nP_n)$ ($\mu_i$ friction coefficients) is a practical notation. The lagrangian components are $J_{\bf q}^T{\bf X}{\cal F}_R=-{\Bbb D}{\dot {\bf q}}$, being ${\Bbb D}({\bf q}, t)=J_{\bf q}^T{\bf X}\,J_{\bf q}{\bf X}_\mu$ symmetric and positive definite, and ${\cal R}=-\frac{1}{2}{\dot {\bf q}}\cdot {\Bbb D}{\dot {\bf q}}$. Condition (\ref{rayr}) writes in this case $\nabla_{\dot {\bf q}}{\cal R}=-{\Bbb D}{\dot {\bf q}}$.

\noindent
With regard to case $(ii)$, the device of defining 
the function $\gamma_\varrho({\bf q}, {\dot {\bf q}}, \dots, \cub^{\varrho \cdot}, t)=J_{\bf q}^T{\bf X}{\cal G}\cdot \cub^{\varrho\cdot}$ provides, according to (\ref{point}), 
\begin{equation}
\label{forzer}
{\cal O}_\varrho[\gamma_\varrho] = J_{\bf q}^T{\bf X}{\cal G}.
\end{equation}
In this way the lagrangian components of a generic force ${\cal G}
({\bf q}, \dots, \cub^{(\varrho-1)\cdot}, t)$ can be intercepted by the operator
(\ref{extlagr}) by means of a function $\gamma_\varrho$, at the expense of expanding the variables to the subsequent derivative $\cub^{\varrho\cdot}$. 

\section{Equations of motion}

\noindent
We are going now to assemble the various aspects: the main objective is to encompass contributions of different kinds under the one equation (\ref{extlagr}), for appropriate $r$ and ${\cal Y}_r$. 

\noindent
Basing on the Newton's law splitted on the lagrangian components
\begin{equation}
\label{newton}
J_{\bf q}^T{\bf X}{\dot \cu}=J_{\bf q}^T{\bf X}{\cal F},
\end{equation}
we model the dynamical term ${\cal F}$. 
We separate the contributions according to 
\begin{equation}
\label{f}
{\cal F}={\cal F}_0+{\cal F}_S+{\cal F}_A+{\cal F}_R+{\cal G}
\end{equation}
exhibiting the following features:

\begin{itemize}
\item[$(i)$] ${\cal F}_0$ originates from a potential $U_0({\bf q},t)$ such that 
$J_{\bf q}^T{\bf X}{\cal F}_0=\nabla_{\bf q}U_0$, 
\item[$(ii)$] the term ${\cal F}_S$ corresponds to a generalized potential and verifies (\ref{potgen}) for some $s\geq 0$, so that $J_{\bf q}^T{\bf X}{\cal F}_S={\cal O}_s[{\cal U}_s]$ for an appropriate ${\cal U}_s$,  

\item[$(iii)$] the contribution ${\cal F}_A$ is caused by acceleration effects or by higher order effects and it accounts for condition (\ref{fa}) with a suitable integer $h\geq 1$; the function (\ref{ul}) provides the lagrangian components of the force according to ${\cal O}_{h-1}[{\cal U}_A]=-hJ_{\bf q}^T{\bf X} {\cal F}_A$;

\item[$(iv)$] the term ${\cal F}_R$ is of the (\ref{rayr}) type for some $\sigma \geq 1$ and the lagrangian components verify (\ref{complagrfr}), 
\item[$(v)$] the contribution ${\cal G}$ concerns a generic force depending on ${\bf q}$, ${\dot {\bf q}}$, $\dots$, $\cub^{(\varrho-1)\cdot}$ and possibly $t$ for some $\varrho \geq 1$: the formal settlement to have in mind is (\ref{forzer}).
\end{itemize}

\noindent
For a given system of forces (\ref{f}) we set $r=\max (s,m-1,\sigma, \varrho)$.
On the ground of the previous analysis, we are going to prove the following

\begin{prop}
The equations of motion for the set of material points $(m_i,P_i)$, $i=1,\dots, N$ subject to the system of forces (\ref{f}) are 
\begin{equation}
\label{eqmoto}
{\cal O}_r[{\cal Y}_r]=\nabla_{\cub^{r\cdot}} {\cal Y}_r
-(r+1)\frac{d}{dt}\nabla_{\cub^{(r+1)\cdot}} {\cal Y}_r=0
\end{equation}
where
\begin{equation}
\label{yr}
{\cal Y}_r = \frac{d^r{\cal L}}{dt^r}
+\frac{d^{r-s}U_S}{dt^{r-s}}
-\frac{1}{h}\dfrac{d^{r-(h-1)}U_A}{dt^{r-(h-1)}}
-\frac{d^{r-\sigma}{\cal R}}{dt^{r-\sigma}}
+\frac{d^{r-\varrho}\gamma_\varrho}{dt^{r-\varrho}}
\end{equation}
and ${\cal L}=\frac{1}{2}\cu\cdot {\dot {\bf X}}+U_0$, where ${\bf Q}$ is the $3N$--vector of the momenta $m_i {\dot P}_i$. 
\end{prop}

\noindent
{\bf Proof}: By virtue of Propositions 1.2, 1.3 and the linearity of (\ref{extlagr}) we can write
$$
{\cal O}_r[{\cal Y}_r]={\cal O}_0[{\cal L}] +{\cal O}_s[{\cal U}_S]-\frac{1}{m} {\cal O}_{m-1}[{\cal U}_A] - {\cal O}_\sigma [{\cal R}] +{\cal O}_\varrho[\gamma_\varrho].
$$
On the one hand, the first of (\ref{tuop}) calculated for $r=0$ means ${\cal O}_0[{\cal L}]=-J_{\bf q}^T{\bf X}{\dot {\bf Q}}+\nabla_{\bf q}U_0$; on the other hand, the second in (\ref{tuop}) and (\ref{ul}), (\ref{complagrfr}), (\ref{forzer}) entail
$$
{\cal O}_s[{\cal U}_S]-\frac{1}{m} {\cal O}_{m-1}[{\cal U}_A] - {\cal O}_\sigma [{\cal R}] +{\cal O}_\varrho[\gamma_\varrho]=J_{\bf q}^T{\bf X}({\cal F}_S+{\cal F}_A+{\cal F}_R+{\cal G})
$$
so that (\ref{eqmoto}) is equivalent to $J_{\bf q}^T{\bf X}(-{\dot {\bf Q}}+{\cal F})=0$, that is (\ref{newton}). $\quad\square$

\noindent
In the end, we remark that combining (\ref{extlagr}) with (\ref{prop2}) calculated for $n=r+1$ and $P={\cal Y}_r$, one finds
\begin{equation}
\label{niels0}
{\cal O}_r[{\cal Y}_r] = \nabla_{\cub^{(r+1)\cdot}}\frac{d{\cal Y}_r}{dt}-(r+2)
\frac{d}{dt}\nabla_{\cub^{(r+1)\cdot}} {\cal Y}_r.
\end{equation}
Whenever ${\cal O}_r[{\cal Y}_r]=0$ gives the equations of motion, by replacing $\frac{d}{dt}\nabla_{\cub^{(r+1)\cdot}} {\cal Y}_r$ with $\frac{1}{r+1}\nabla_{\cub^{r\cdot}} {\cal Y}_r$
in (\ref{niels0}) one achieves
\begin{equation}
\label{nielsen}
\nabla_{\cub^{(r+1)\cdot}}\frac{d{\cal Y}_r}{dt}-\frac{r+2}{r+1}
\nabla_{\cub^{r\cdot}} {\cal Y}_r=0
\end{equation}
which can be considered a sort of \emph{generalized Nielsen's equations}, in the sense that 
the case $r=0$
\begin{equation}
\label{nielsen}
\nabla_{\dot \cub}\frac{d{\cal L}}{dt}-2\nabla_{\cub} {\cal L}=0
\end{equation}
for ${\cal L}({\bf q}, {\dot {\bf q}}, t)$ 
corresponds to the set known as the \emph{Nielsen form of the equations of motion} (a relatively recent study is in \cite{niels}).

\section{Conclusions}

\noindent
The covariance with respect to changes of coordinates of the binomial expression (\ref{extlagr}) provides access to make use of it for the motion of a systems under wide assumtions about the dynamical strains.
The physical motivation of the higher order operator passes through the Newton's law and the lagrangian components of an overall force (\ref{f}) including generalized potentials (\ref{potgen}), kinematic effects (\ref{fa}), generalized dissipative forces (\ref{rayr}) or even non structured forces.
For specific values of the indexes, the standard equations for the known and ordinary cases are reproduced.

\noindent
By means of the property (\ref{prop13}), all these effects can be encompassed by the the function ${\cal Y}_r$, where $r$ depends on the highest order of derivative occurring in each contribution.
The order of the equations of motion (\ref{eqmoto}) is $r+2$. 
As (\ref{yr}) shows, the unifying procedure possibly requires derivations on the individual terms: on the other hand, the compact expression (\ref{eqmoto}) may facilitate the investigation on specific properties of the system, in the same way as the energy balance can be deduced from the ordinary lagrangian equations of motion.

\noindent
At the same time it is interesting to detect that some alternative form of the equations of motion, involving the derivatives of state functions (Nielsen or Appell equations) can be recovered from the set of equations (\ref{eqmoto}).

\noindent
An interesting further development will be the investigation of the expressions (\ref{opercovar}), for $h\not =1$, especially with regard to their bearing on real physical systems.
On the other hand, the context is well suited for letting more the constraint equations more general, embracing linear and nonlinear holonomic restrictions, in order to develop and complete the starting results formulated in \cite{tal}.

\end{document}